\documentclass[%
 reprint,
superscriptaddress,
 amsmath,amssymb,
 aps,
pre,
floatfix,
]{revtex4-2}
\usepackage{booktabs}
\usepackage{multirow}
\usepackage[colorlinks=true,urlcolor=blue,linkcolor=blue,citecolor=blue]{hyperref}
\usepackage[normalem]{ulem}
\newcommand{\customref}[2]{\hyperref[#1]{\ref*{#1}#2}}
\usepackage{graphicx}
\usepackage{bm}
\usepackage{hyperref}
\usepackage{color}
\usepackage[table,dvipsnames]{xcolor} 
\usepackage{makecell} 
\usepackage{booktabs}
\usepackage{float}

\usepackage{lineno}

\usepackage{lipsum}
\begin{document}

\preprint{APS/123-QED}

\title{Reusability Report: Optimizing T-count in General Quantum Circuits with AlphaTensor-Quantum}

\author{Remmy Zen}
\email{remmy.zen@mpl.mpg.de}
\affiliation{Max Planck Institute for the Science of Light, Staudtstra{\ss}e 2, 91058 Erlangen, Germany}
\author{Maximilian N\"{a}gele}
\email{maximilian.naegele@mpl.mpg.de}
\affiliation{Max Planck Institute for the Science of Light, Staudtstra{\ss}e 2, 91058 Erlangen, Germany}
\affiliation{Department of Physics, Friedrich-Alexander Universit\"{a}t Erlangen-N\"{u}rnberg, Staudtstra{\ss}e 5, 91058 Erlangen, Germany}
\author{Florian Marquardt}%
\email{florian.marquardt@mpl.mpg.de}
\affiliation{Max Planck Institute for the Science of Light, Staudtstra{\ss}e 2, 91058 Erlangen, Germany}
\affiliation{Department of Physics, Friedrich-Alexander Universit\"{a}t Erlangen-N\"{u}rnberg, Staudtstra{\ss}e 5, 91058 Erlangen, Germany}

\date{\today}

\begin{abstract}
Quantum computing has the potential to solve problems that are intractable for classical computers, with possible applications in areas such as drug discovery and high-energy physics. However, the practical implementation of quantum computation is hindered by the complexity of executing quantum circuits on hardware. In particular, minimizing the number of T-gates is crucial for implementing efficient quantum algorithms.
AlphaTensor-Quantum~\cite{ruiz_quantum_2024} is a reinforcement learning-based method designed to optimize the T-count of quantum circuits by formulating the problem as a tensor decomposition task. While it has demonstrated superior performance over existing methods on benchmark quantum arithmetic circuits, its applicability has so far been restricted to specific circuit families, requiring separate, time-intensive training for each new application.
This report reproduces some of the key results of the original work and extends AlphaTensor-Quantum's capabilities to simplify random quantum circuits with varying qubit counts, eliminating the need for retraining on new circuits. Our experiments show that a general agent trained on 5- to 8-qubit circuits achieves greater T-count reduction than previous methods for a large fraction of quantum circuits. Furthermore, we demonstrate that a general agent trained on circuits with varying qubit numbers outperforms agents trained on fixed qubit numbers, highlighting the method's generalizability and its potential for broader quantum circuit optimization tasks.
\end{abstract}

\maketitle

Reinforcement learning (RL)~\cite{sutton2018reinforcement} is a framework for discovering optimal action sequences in decision-making problems where the best strategy is unknown and often nontrivial. In recent years, deep RL has transformed problem-solving across multiple fields, such as robotics~\cite{tang_deep_2024}, drug discovery~\cite{zhou_optimization_2019}, and game-playing, where AlphaZero~\cite{silver_mastering_2017} surpassed human experts in board games such as Go and Chess. This approach was later extended to tackle problems in the field of mathematics, where AlphaZero was adapted to discover a more efficient and provably correct algorithm for matrix multiplication~\cite{fawzi_discovering_2022}. The resulting agent, AlphaTensor, was trained to play a TensorGame with the goal of finding efficient tensor decompositions.

Quantum computation is an emerging technology promising exponential speedups over classical computation for certain problems such as cryptography~\cite{shor_polynomial-time_1997} and quantum simulation~\cite{daley_practical_2022}.  This has potentially extensive implications, from securing communications~\cite{kimble_quantum_2008} to advancing drug discovery~\cite{2022_perspective}.  However, a major bottleneck in practical quantum computing is the complexity of the quantum circuits required to implement quantum algorithms. In particular, the T-gate, a fundamental quantum logic gate, is one of the most resource-intensive to implement~\cite{campbell_roads_2017,beverland_assessing_2022}. Despite this, T-gates are essential for achieving universal quantum computation~\cite{nielsen2010quantum}. Therefore, reducing the T-count of quantum circuits is crucial before implementing them on quantum hardware.

Several methods have been developed for optimizing quantum circuits~\cite{2020_reducing,heyfron_efficient_2018,amy_polynomial-time_2014,yamashita_reversible_2014}, including machine learning~\cite{daimon_quantum_2024} and reinforcement learning techniques~\cite{fosel_quantum_2021,li_quarl_2023,riu_reinforcement_2024}. More recently, AlphaTensor-Quantum~\cite{ruiz_quantum_2024} extended AlphaTensor’s capabilities into the field of quantum computing by formulating T-count optimization as a tensor decomposition problem. Unlike AlphaTensor, AlphaTensor-Quantum can incorporate domain-specific knowledge by using gadgets, which is a procedure to reduce T gates by using ancillary qubits, to enhance optimization efficiency. Additionally, AlphaTensor-Quantum introduces symmetrized axial attention layers in its neural network, which take advantage of the signature tensor's symmetry, thereby allowing it to scale to larger qubit numbers.

On a benchmark of quantum arithmetic circuits, AlphaTensor-Quantum has been shown to achieve a lower T-count than previous existing methods, particularly when gadgets are incorporated. However, its training is limited to specific quantum circuits grouped by application. This means that the model must be retrained for each new type of application, resulting in increased computational cost. In this paper, we first evaluate the reproducibility of AlphaTensor-Quantum’s results. We then extend its application to a more general quantum circuit optimization problem: training a single agent capable of optimizing random quantum circuits with varying numbers of qubits and gates. This approach enables faster optimization without the need for retraining on each new circuit. The general agent achieves a lower T-count on a large fraction of circuits compared to the baseline and the agents trained on fixed qubit sizes.

\section*{Reproducibility} 

\begin{figure}[htb]
	\includegraphics[width=.47\textwidth]{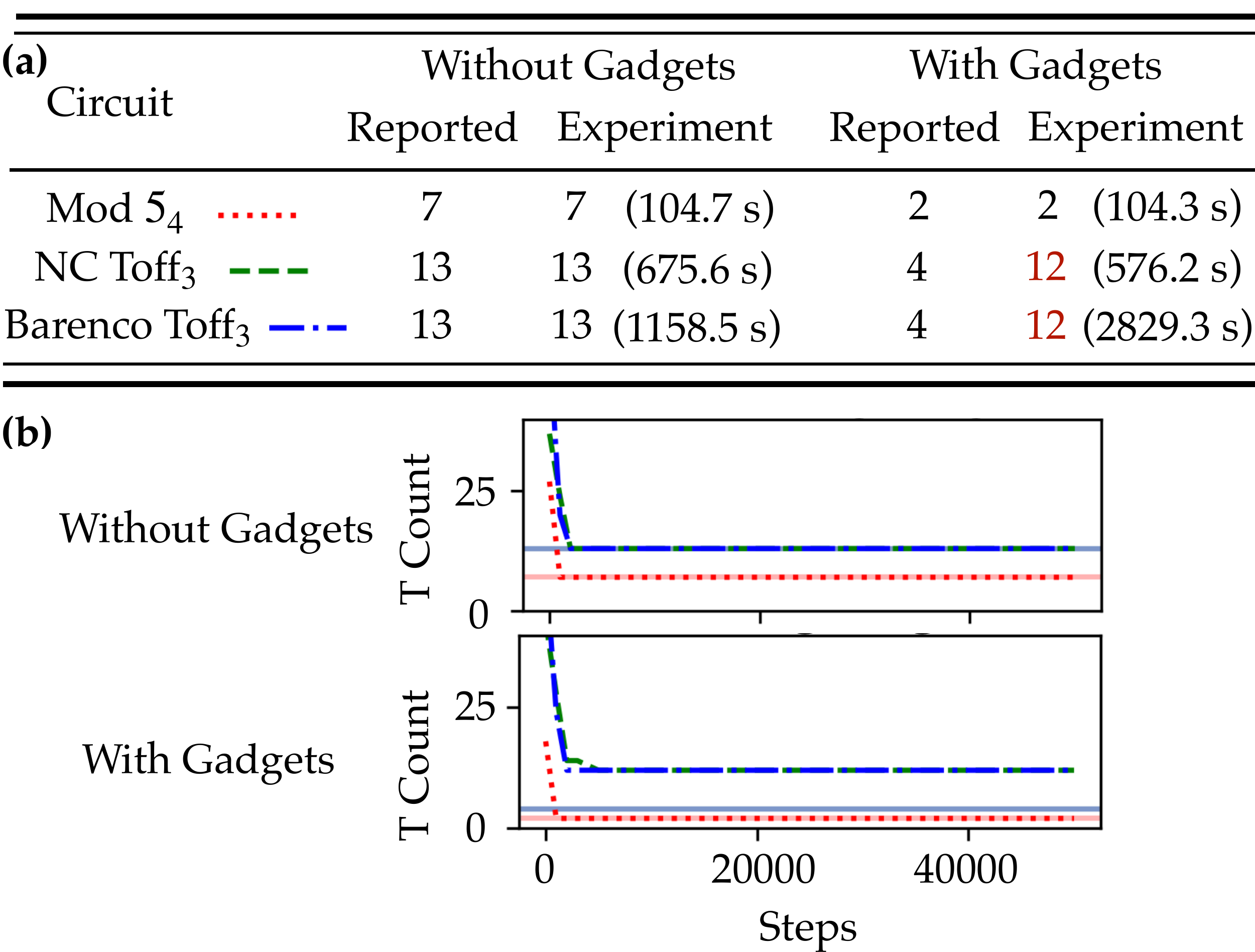}
	\caption{ \label{fig:reproducibility} Reproducing AlphaTensor-Quantum. \textbf{(a)}   The T-count reported in the original manuscript along with the results from experiments using the provided code. The training time to reach optimal performance on an NVIDIA A100 GPU is given in parentheses. Red numbers indicate where our experimental results do not match the originally reported values (see main text). \textbf{(b)} The evolution of the T-count during training. The light solid lines represent the reported result.}
\end{figure}

The original publication on AlphaTensor-Quantum relies on AlphaTensor, which in turn builds on AlphaZero. Unfortunately, neither the code for AlphaZero nor AlphaTensor has been made publicly available at this time. In addition, the in-house computing resources and infrastructure being used at Google DeepMind for this project are beyond the scale and sophistication of what is available in an academic context.

Nevertheless, the authors have made (slightly revised) {\em parts} of their code available in the GitHub repository~\cite{githubrepo}. This repository includes implementations of the TensorGame and their neural network architecture, integrated into the publicly available Monte Carlo Tree Search (MCTS) framework MCTX~\cite{deepmind2020jax} (as a replacement for AlphaZero). Again, we emphasize that the results presented in the paper were not obtained using this specific MCTS framework, and the implementation details differ, which may explain some of the discrepancies observed in our numerical experiments.

In addition to the code, the signature tensors for the circuits Mod $5_4$ (5 qubits), NC Toff$_3$ (7 qubits), and Barenco Toff$_3$ (8 qubits) from Table 2 in~\cite{ruiz_quantum_2024} are provided for testing in the GitHub repository. In the following, we aim to reproduce the results for these three circuits. Reproducing other findings proved challenging, as the authors of~\cite{ruiz_quantum_2024} do not provide code to generate signature tensors from a quantum circuit and do not specify the exact hyperparameters used for each experiment.  See Methods for discussion about the hyperparameters. 
We present the optimized T-count and training time in Fig.~\ref{fig:reproducibility}(a). We observe that the T-count for NC Toff$_3$ and Barenco Toff$_3$ with gadgets is higher than originally reported. By doubling the batch size and the number of MCTS simulations, the T-count is reduced to 8 for NC Toff$_3$ and 10 for Barenco Toff$_3$, suggesting that further hyperparameter tuning could likely reproduce the original findings. It is also worth noting that in the original paper, AlphaZero-Quantum is trained on a family of circuit applications. For example, for the Barenco Toffoli application, AlphaZero-Quantum is trained on Barenco Toff$_3$, Barenco Toff$_4$, Barenco Toff$_5$, and Barenco Toff$_{10}$ circuits. By contrast, in our case, the optimization of Barenco Toff$_3$ is trained only based on the Barenco Toff$_3$ circuit, because the circuits or the tensor representation of the circuits are not available.
Fig.\ref{fig:reproducibility}(b) shows the evolution of the T-count throughout training. The T-count converges after approximately 3000 training steps, which takes between $100$ and $1000$ seconds depending on the number of qubits. Additionally, we train a single agent to simplify all provided circuits simultaneously, achieving the same performance (see Supplementary Figure 1).

\begin{figure}[tb]
	\includegraphics[width=.4\textwidth]{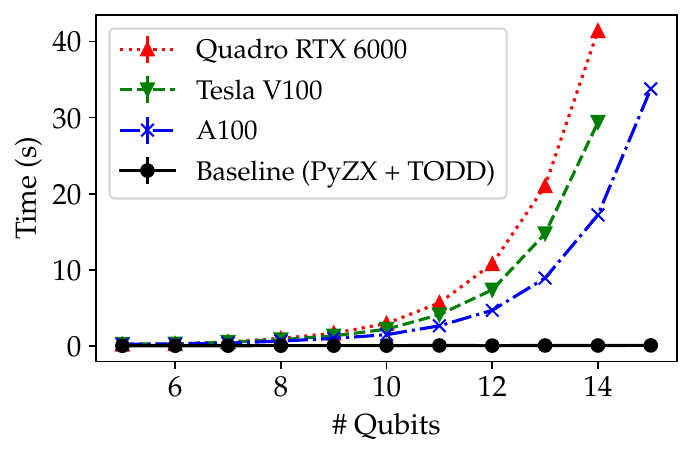}
	\caption{ \label{fig:time-scale} Average time for one step of AlphaTensor-Quantum training with   gadgets on different GPU devices. Quadro RTX 6000 and Tesla V100 give an out-of-memory error for $15$ qubits. We compare with the baseline PyZX~\cite{2020_reducing} and TODD~\cite{heyfron_efficient_2018}. Note that PyZX and TODD directly output the optimized circuit in the given time (e.g. around 0.06 seconds for 15 qubits). By contrast, AlphaTensor-Quantum requires a large number of training steps (e.g.\ between tens of thousands and several million in the original manuscript).  Error bars, corresponding to one standard deviation across $10$ different circuits, are smaller than the marker size.}
\end{figure}

\begin{figure*}[!htb]
	\includegraphics[width=.85\textwidth]{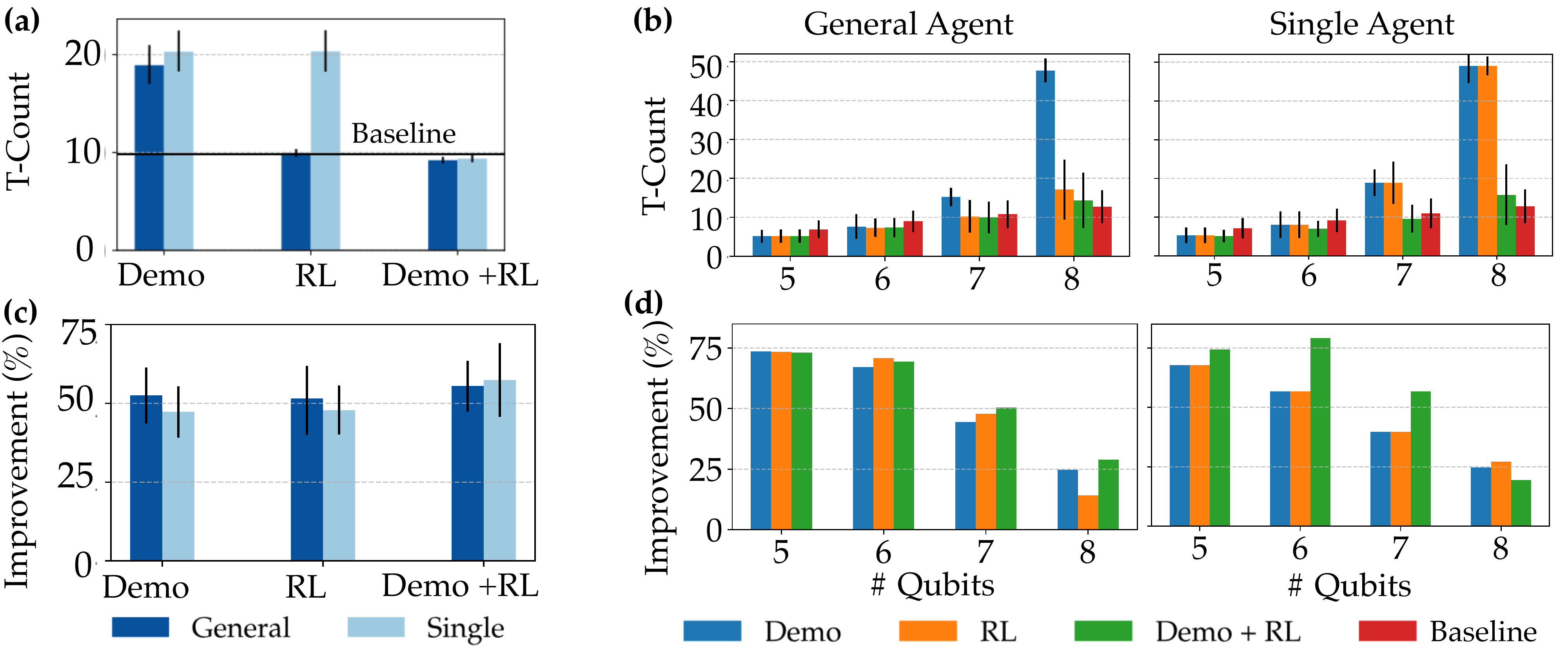}
	\caption{ \label{fig:results} Evaluation of single (random circuits, fixed qubit number) and general (random circuits, varying qubit number) AlphaTensor-Quantum agents with gadgetization and three training types: only with synthetic demonstrations (Demo), only with reinforcement learning (RL), and both (Demo + RL). \textbf{(a)} The average T-count (lower is better) of the optimized quantum circuits in the evaluation set.  The solid black line shows the average T-count of the baseline method PyZX~\cite{2020_reducing} and TODD~\cite{heyfron_efficient_2018}. \textbf{(b)} The average T-count for each number of qubits.  \textbf{(c)} The average improvement percentage (higher is better), which shows the percentage of circuits that have a strictly lower T-count when optimized with the agent compared to the baseline method. \textbf{(d)}  The average improvement percentage for each number of qubits. The error bar for \textbf{(a)} and \textbf{(c)} shows the 95\% confidence intervals over different numbers of qubits and for \textbf{(b)} shows the 95\% confidence intervals over $1000$ evaluation circuits. }
\end{figure*}

Since the provided example tensors correspond to small numbers of qubits, we also examine the expected runtime of AlphaTensor-Quantum for larger circuits. In Fig.~\ref{fig:time-scale}, we illustrate how the training time of AlphaTensor-Quantum scales with the number of qubits on different GPU devices using the provided hyperparameters. In this case, we fixed the task to optimize a random circuit where the number of gates is fixed to be $10$ times the number of qubits and half of them are $T$ gates.
We observe that the training time for AlphaTensor-Quantum increases exponentially. This is likely due to the exponentially increasing number of possible actions for AlphaTensor-Quantum. In the original manuscript~\cite{ruiz_quantum_2024}, the amount of sampled actions is therefore kept to a fixed maximum number, which is not implemented in the provided code.
The baseline method using PyZX~\cite{kissinger_pyzx_2020, 2020_reducing} and TODD~\cite{heyfron_efficient_2018} is several orders of magnitude faster than AlphaTensor-Quantum which  requires training for tens of thousands to several million steps per optimized circuit. Consequently, the computational overhead of running AlphaTensor-Quantum is likely justified only for important quantum circuit primitives that serve as building blocks for numerous applications.

\section*{Generalizability}

To improve the optimization efficiency of AlphaTensor-Quantum by eliminating the need for retraining on previously unseen circuits, we train it to simplify random quantum circuits spanning multiple qubit sizes. We refer to this agent as the \textbf{general agent}. We then compare its performance to agents trained separately for each qubit size. We refer to these agents as the \textbf{single agents}. Note that the single agents are already more general than the AlphaTensor-Quantum agents used in~\cite{ruiz_quantum_2024}, which are trained on specific quantum circuit applications.

In our experiments, we use quantum circuits with 5 to 8 qubits.  Therefore, we train one general agent across all these qubit numbers and four separate single agents, each for a specific qubit number. AlphaTensor-Quantum is originally trained using a combination of supervised learning on synthetic demonstrations and reinforcement learning on the target circuits. The synthetic demonstrations dataset consists of randomly generated tensor/factorization pairs for the neural network to imitate.
To evaluate the contribution of these components,
we train our agents either only with synthetic demonstration data (\textbf{Demo}), only with reinforcement learning data (\textbf{RL}), or with both (\textbf{Demo + RL}). See Methods for details of the dataset generation and training process. For RL and Demo + RL, we use $100000$ random circuits for reinforcement learning. We first focus on the AlphaTensor-Quantum version that includes gadgetization (see Supplementary Figure 2 for results without gadgetization).  We train AlphaTensor-Quantum with the default hyperparameters for $100000$ steps. For each considered qubit number, we generate $1000$ random quantum circuits as the evaluation set. During the evaluation, we always choose the most probable action predicted by the MCTS policy. As a baseline, we optimize these circuits with PyZX~\cite{2020_reducing} and then apply TODD~\cite{heyfron_efficient_2018}, as done in~\cite{ruiz_quantum_2024}.

We first evaluate the average T-count in Fig~\ref{fig:results}(a) for single and general agents trained with the three training types. The general agent consistently outperforms the single agents across all training types. Additionally, the Demo + RL agent achieves the lowest average T-count, falling below the baseline indicating that the mix of supervised demonstration and RL training is useful.

Fig~\ref{fig:results}(b) presents the performance of the agents across different qubit sizes. As expected, the average final T-count grows with the qubit number since the sampled initial circuits contain more T gates. However, the optimization is increasingly less effective compared to the baseline with higher qubit count. Notably, the agents outperform the baseline for $N = 5$ and $6$. However, performance declines at $N = 7$, except when using Demo + RL training, and further deteriorates at $N = 8$, where all agents perform worse than the baseline on average with only about 23\% improvement. The performance could be improved by hyperparameter tuning and longer training.

Although Fig.\ref{fig:results}(a) and (b) demonstrate average T-count reduction, this alone does not fully capture how consistently the agents outperform the baseline. To address this issue, we introduce the \textbf{improvement percentage} metric, which measures the fraction of circuits in the evaluation set where the agent achieves a strictly lower T-count than the baseline. It is important to note that the input to the AlphaTensor-Quantum is already a circuit that is optimized with PyZX following the compilation method described in~\cite{ruiz_quantum_2024}.

Fig.\ref{fig:results}(c) shows that all agents in general achieve an improvement above 45\% compared to the baseline, with Demo + RL again outperforming the other training types. The general agent surpasses the single agent overall, except with the Demo + RL training type. Fig.\ref{fig:results}(d) further confirms the trend discussed before that the improvement percentage declines as circuit size increases. A significant improvement is observed for $N = 5$ and $6$, while it diminishes for $N = 7$ and $8$. A similar trend is observed for AlphaTensor-Quantum without gadgetization (see Supplementary Figure 2).

\begin{figure}[tb]
	\includegraphics[width=.3\textwidth]{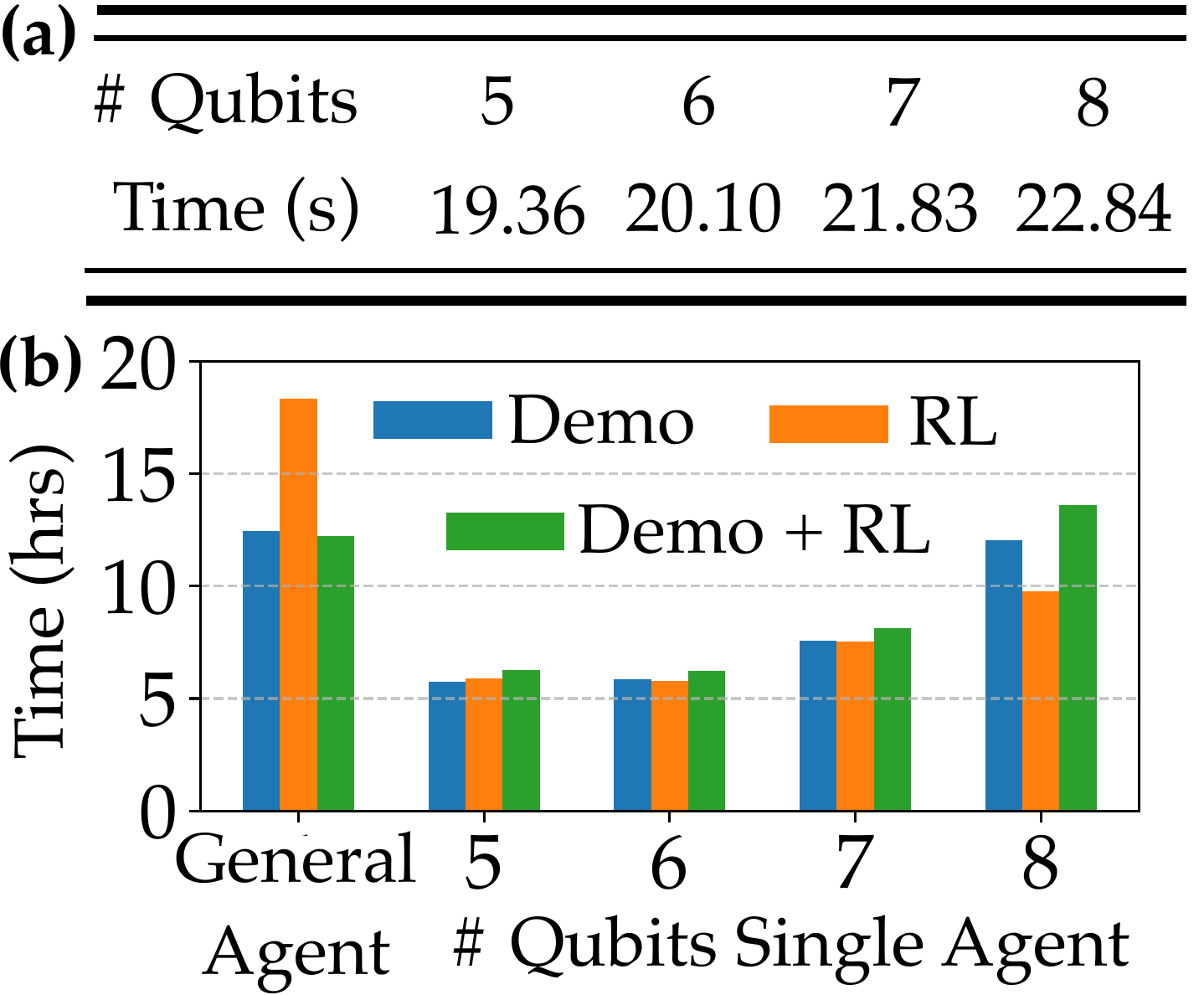}
	\caption{ \label{fig:time-plot}
 Average time required on a single NVIDIA A100 GPU to \textbf{(a)} simplify a single circuit during evaluation
 and \textbf{(b)} to train the agents for 100000 steps. }

\end{figure}

A key advantage of our agents is their fast execution time during evaluation. Unlike the original AlphaTensor-Quantum, which requires retraining for each circuit, a process that can take from a few minutes to several hours, our pre-trained agents simplify circuits in a single rollout, averaging around 20 seconds (see Fig.~\ref{fig:time-plot}(a)).

\begin{table}[bh]
    \centering
    \caption{T-count after optimizing benchmark circuits with the general agent trained with different training methods. The bold numbers indicate the number matches with the reported values.
 \label{tab:known_circuits}}

    \renewcommand{\arraystretch}{1.2} 
    \setlength{\tabcolsep}{4pt} 
    \begin{tabular}{@{}l l c c c@{}}
        \toprule        
        \midrule

        & & Mod $5_4$ & NC Toff$_3$ & Barenco Toff$_3$ \\ 
        \midrule
        \multirow{4}{*}{\makecell[l]{{Without} \\ {Gadgets}}} 
        & Reported   & 7  & 13 & 13  \\ 
        & Demo      & \textbf{7}  & \textbf{13} & \textbf{13}  \\ 
        & RL        & \textbf{7}  & 14 & 14  \\ 
        & Demo + RL & 8  & 14 & 14  \\ 
        \midrule
        \multirow{4}{*}{\makecell[l]{{With} \\ {Gadgets}}}  
        & Reported   & 2  & 4  & 4   \\ 
        & Demo      & \textbf{2}  & \textbf{4}  & \textbf{4}   \\ 
        & RL        & 7  & 15 & 13  \\ 
        & Demo + RL & 6  & 14 & 12  \\ 
                \midrule
        \bottomrule
    \end{tabular}
\end{table}

Finally, in Table~\ref{tab:known_circuits}, we evaluate our general agents on the three target circuits in Fig.~\ref{fig:reproducibility}, which the agents never encountered before during training. The Demo agent finds the optimal T-count both with and without gadgets, while the other two methods perform slightly worse.

\section*{Conclusion and Discussion}

In this work, we first assess the reproducibility of AlphaTensor-Quantum~\cite{ruiz_quantum_2024}. We find that reproduction of small scale experiments is feasible, albeit potentially requiring some hyperparameter. We then study the generalizability of AlphaTensor-Quantum for general quantum circuit optimization across different qubit sizes. Our approach eliminates the need for retraining on previously unseen circuits, accelerating the optimization process by orders of magnitude compared to the original AlphaTensor-Quantum approach trained on specific quantum circuit applications. From an application perspective, these agents can be integrated with traditional T-count optimizers to achieve further reductions in a large fraction of circuits.

Our experiments demonstrate that a general agent trained on circuits with varying qubit sizes outperforms single agents specialized for a fixed qubit size, highlighting its ability to generalize effectively to unseen circuits when trained on diverse data. The best results are obtained by combining supervised learning on demonstration data with reinforcement learning. However, even agents trained solely on potentially suboptimal supervised demonstrations prove to be effective.

Note that the results presented in this paper are obtained without hyperparameter tuning and require several orders of magnitude less computation than the original AlphaTensor-Quantum training (e.g.\ $10\times$ more training steps, $10\times$ more simulated trajectories per MCTS step, $16\times$ larger batch size, over 3600 TPUs used). This suggests a promising path for scaling to higher qubit numbers by increasing computational resources.

The code from the GitHub repository~\cite{githubrepo} is well-documented and easy to use. While this code differs from the code used to produce the results in~\cite{ruiz_quantum_2024}, the implementations of the symmetrized axial attention layers and the TensorGame environment provide valuable building blocks for future research. Integration with the MCTX Monte Carlo Tree Search library enables rapid reproduction of some of the results from the original work. The additional GitHub repository~\cite{circuittotensor} provides functionality to compute the signature tensor of a given quantum circuit and implements a post-processing pipeline to reconstruct the optimized circuits, which is crucial for their practical implementation.  However, providing the exact hyperparameters and the exact code used in the original paper would enhance reproducibility and assist in choosing optimal hyperparameters for future work.

Looking ahead, AlphaTensor-Quantum has the potential to serve as a powerful framework for minimizing the T-count of quantum circuit primitives when the computational cost is justified by their importance. Additionally, general agents such as those trained in this manuscript offer a promising middle ground between traditional T-count optimizers and the original AlphaTensor-Quantum approach, balancing computational efficiency and performance.

\section*{Methods}

\subsection*{Hyperparameters for reproducibility experiments}

In our experiments, we use the default hyperparameter given in the GitHub repository. For example, the batch size is 2048 and the number of MCTS simulations is 800 in the original paper, while the GitHub implementation uses 128 for the batch size and 80 for the number of MCTS simulations. We have tried to use the hyperparameters from the original paper, but it gives an out-of-memory error. We run the experiments with the provided hyperparameters on a single NVIDIA A100 GPU with 40GB memory.

\subsection*{Dataset generation and training process for generalizability experiments}

To create the training data, we generate random CNOT+T circuits with random number of qubits $N$, selecting the total number of gates uniformly between $5N$ and $15N$, with T gates comprising between 20\% and 60\% of the total gate count. We then follow the quantum circuit compilation approach outlined in~\cite{ruiz_quantum_2024}, which applies an optimization algorithm in PyZX~\cite{2020_reducing} to reduce the initial T-gate count and extract the signature tensor as an input to AlphaTensor-Quantum. If one is interested in a general circuit, one can follow the
Supplementary Section C.1 of~\cite{ruiz_quantum_2024} to compile a general circuit into a circuit containing only CNOT+T gates.

For RL and Demo + RL, we use $100000$ random circuits for reinforcement learning. We first focus on the AlphaTensor-Quantum version that includes gadgetization (see Supplementary Figure 2 for results without gadgetization).  We train AlphaTensor-Quantum with the default hyperparameters for $100000$ steps. For each considered qubit number, we generate $1000$ random quantum circuits as the evaluation set. During the evaluation, we always choose the most probable action predicted by the MCTS policy. As a baseline, we optimize these circuits with PyZX~\cite{2020_reducing} and then apply TODD~\cite{heyfron_efficient_2018}, as done in~\cite{ruiz_quantum_2024}.

\section*{Data Availability}
The data for reproducing this work are available at~\cite{dataset}.

\section*{Code Availability}
The code for reproducing this work is available at~\cite{code}.

\section*{Acknowledgements}
We thank Jan Olle for fruitful discussions. The research is part of the Munich Quantum Valley,  which is supported by the Bavarian state government with funds from the Hightech Agenda Bayern Plus.

\section*{Author Contributions Statement}
R.Z., M.N., and F.M. conceptualized and designed the study. R.Z and M.N. coded and performed the experiments. R.Z, M.N., and F.M. interpreted the data. All authors wrote and revised the manuscript.

\section*{Competing Interests Statement}
All authors declare no competing interests.

\bibliography{main}

\clearpage

\onecolumngrid

\centering
\huge{Supplementary Material }
\renewcommand{\figurename}{Supplementary Figure}

 \setcounter{figure}{0}
\begin{figure*}[!htb]
	\includegraphics[width=.47\textwidth]{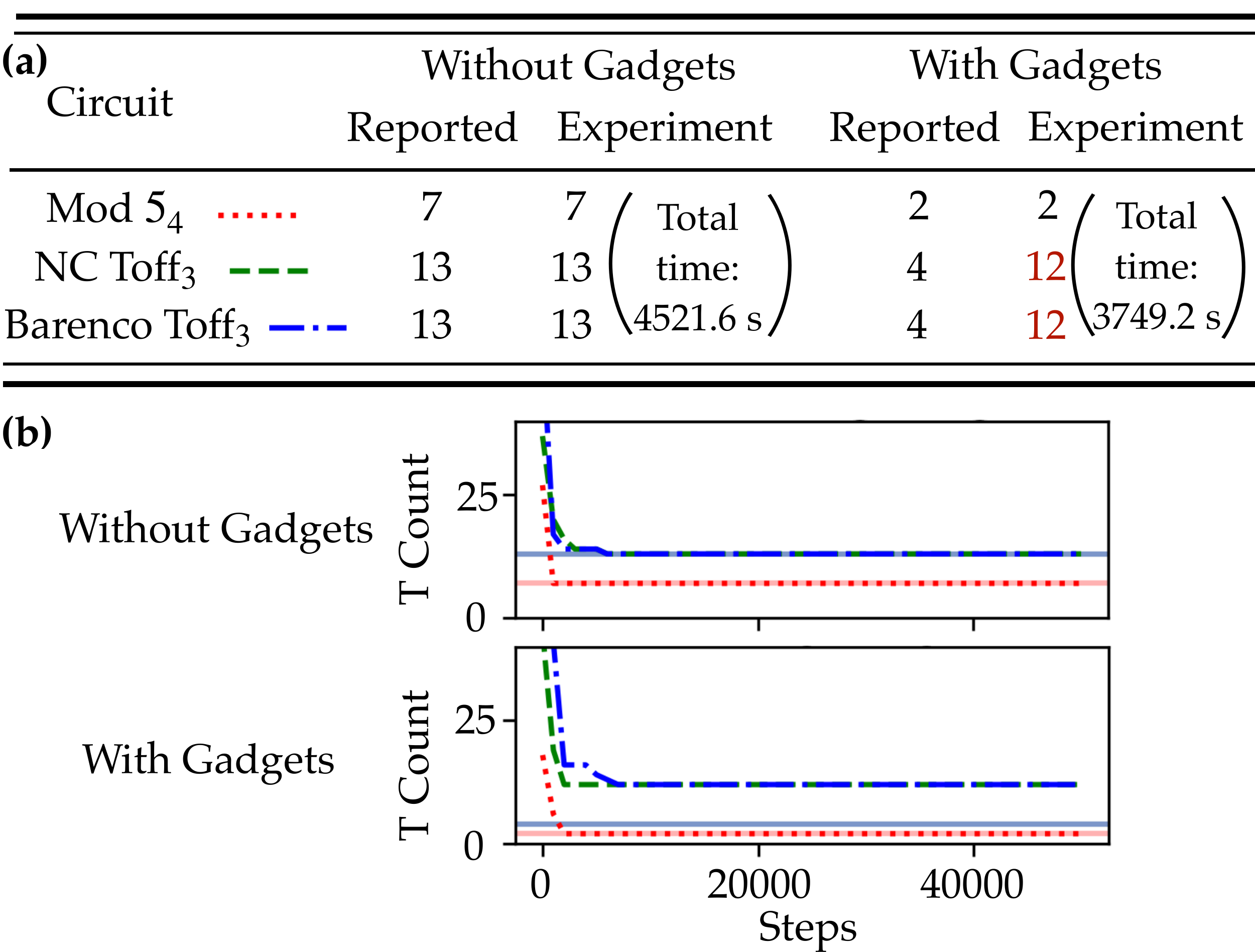}
	\caption{ \label{fig:reproducibility-general} Results for reproducing the AlphaTensor-Quantum by training one agent with the three circuits simultaneously. \textbf{(a)}  The T-count reported in the paper and the results of the experiment with the provided code. The number in brackets indicates the total training time on a NVIDIA A100 GPU. The number in red shows the number where the reported value and the results of the experiments do not agree. \textbf{(b)} The evolution of the T-count during training. The light solid line shows the reported result.}
\end{figure*}

\begin{figure*}[!htb]
	\includegraphics[width=.85\textwidth]{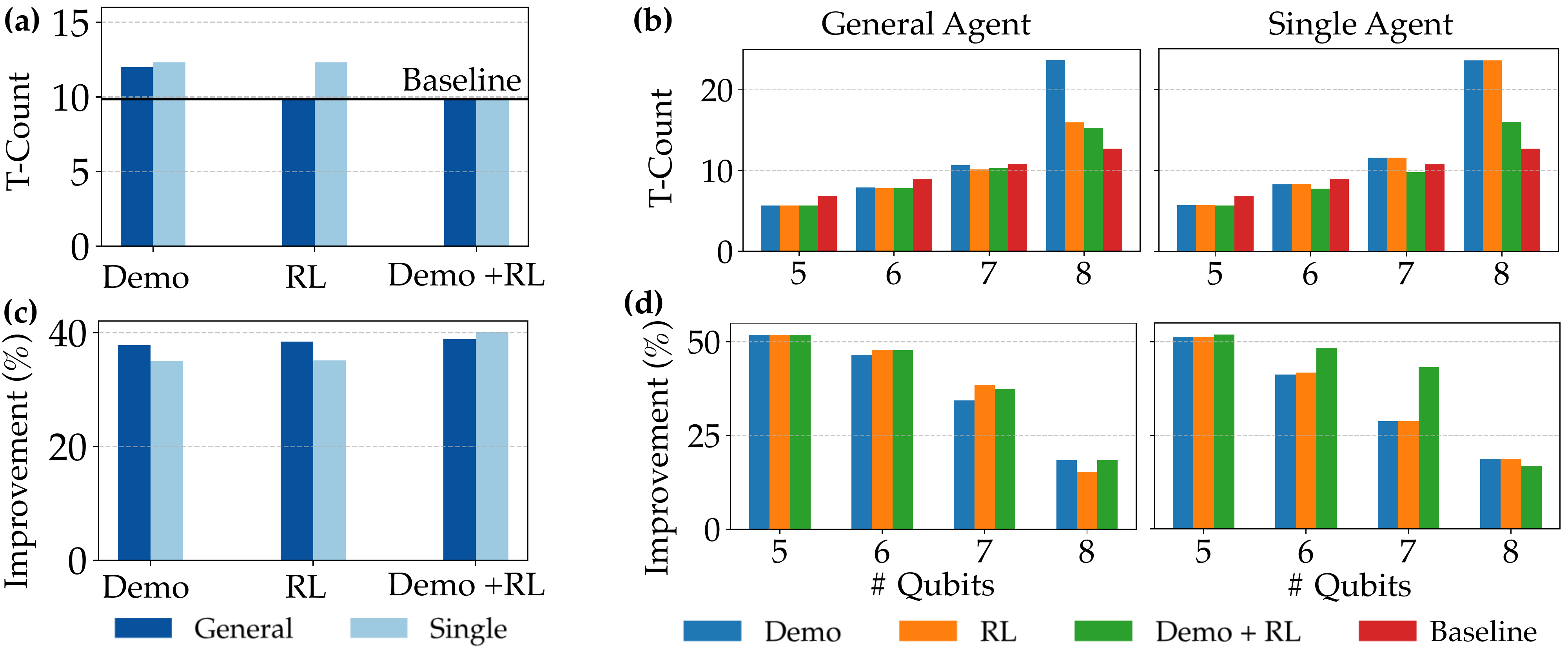}
	\caption{ \label{fig:results-no-gadget} Evaluation of single and general AlphaTensor-Quantum agents without gadgetization and three training types: only with synthetic demonstrations (Demo), only with reinforcement learning (RL), and both (Demo + RL). \textbf{(a)} The average T-count (lower is better) of the optimized quantum circuits in the evaluation set.  The solid black line shows the average T-count of the baseline method PyZX~\cite{2020_reducing} and TODD~\cite{heyfron_efficient_2018}. \textbf{(b)} The average T-count for each number of qubits.  \textbf{(c)} The average improvement percentage (higher is better), which shows the percentage of circuits that have a strictly lower T-count when optimized with the agent compared to the baseline method. \textbf{(d)}  The average improvement percentage for each number of qubits.  }
\end{figure*}

\end{document}